\documentstyle[twocolumn,seceq]{jpsj}
\def\v#1{\mib #1}

\def\runtitle{Ground State of Heisenberg Model with Dimerization and Quadrumerization}
\def\runauthor{Wei {\sc Chen} and Kazuo {\sc Hida}}

\title
{
Ground State Properties of One Dimensional $S=1/2$ Heisenberg Model with Dimerization and Quadrumerization
}

\author
{ 
Wei {\sc Chen}\footnote{chenwei@riron.ged.saitama-u.ac.jp}
and Kazuo {\sc Hida}\footnote{hida@riron.ged.saitama-u.ac.jp}
}

\inst
{
Department of Physics, Saitama University, Urawa, Saitama 338-0825
}

\recdate
{
\today
}

\abst
{
The one dimensional $S=1/2$ Heisenberg model with dimerization and quadrumerization is studied by means of the numerical exact diagonalization of finite size systems. Using the phenomenological renormalization group and finite size scaling law, the ground state phase diagram is obtained in the isotropic case. It exhibits a variety of the ground states which contains the  $S=1$ Haldane state, $S=1$ dimer state and $S=1/2$ dimer state as limiting cases. The gap exponent $\nu$ is also calculated which coincides with the value for the dimerization transition of the isotropic Heisenberg chain. In the $XY$ limit, the phase diagram is obtained analytically and the comparison is made with the isotropic case. 
}

\kword
{
Heisenberg chain, dimerization, quadrumerization, spin gap, exact diagonalization, phenomenological renormalization group, ground state phase diagram, dimer phase, Haldane phase 
}

\begin{document}
\sloppy
\maketitle

\section{Introduction}
Recently, the $S=1/2$ antiferromagnetic Heisenberg chains (AFHC) with  modulated spatial structures have attracted a great deal of attention. Although the uniform $S=1/2$ AFHC can be solved exactly using the Bethe Ansatz method and is known to have a gapless ground state \cite {bh}, this state is unstable against the dimerization leading to the spin-Peierls state which has the spin gap\cite{cf,be,nf,sf}. Other kinds of spatial modulation of the exchange coupling arising from the underlying lattice structures can also induce various kinds of spin gap phases. 

On the other hand,  no spin-Peierls instability is expected in the  $S=1$ AFHC due to the presence of the Haldane gap\cite{hal1}. Instead, the Haldane-dimer phase transition takes place at finite strength of dimerization\cite{tk,kno}.  One of the authors (KH) has pointed out the connection between the dimer state of the $S=1/2$ dimerized Heisenberg chain and the Haldane state of the $S=1$ AFHC\cite{kh1}. In this context, the dimerization in the $S=1$ chain corresponds to the quadrumerization in the dimerized  $S=1/2$ chain.  From this point of view, in the present work, we aim to get deeper insight into the nature of the Haldane-dimer transition in the $S=1$ AFHC from the investigation of the ground state properties of the $S=1/2$ Heisenberg chain with dimerization and quadrumerization. 

On the other hand, in the neighbourhood of the $S=1/2$ uniform Heisenberg point, both  dimerization and quadrumerization are expected to produce the energy gap. It is possible, however, that the coexistence of these two periodicities does not always enhance the gap additively but may rather {\it reduce } the gap due to the competition between them leading to the gapless state in spite of the  spatial nonuniformity. We also intend to investigate  such competition of two periodicities in this model.

This paper is organized as follows. In the next section, the model Hamiltonian is defined. The numerical results are presented in {\S}3 for the isotropic case.  From the phenomenological renormalization group and finite size scaling analysis of the numerically calculated energy gaps, the ground state phase diagram is obtained in \S 3. The critical exponent $\nu$ of the energy gap is also estimated taking the logarithmic corrections into account. The numerical results are compared with the perturbation theory.  In  \S 4, the ground state phase diagram is obtained analytically for the $XY$ case using the  Jordan-Wigner transformation. The comparison is made with the isotropic case. The last section is devoted to summary and discussion.

\section{Model Hamiltonian}
The Hamiltonian of the one dimensional dimerized and quadrumerized $S=1/2$ Heisenberg chain is given by 
\begin{eqnarray} 
{\cal H} &=& \sum_{l=1}^{2N} \Big[ j(S_{2l-1}^{x}S_{2l}^{x}+S_{2l-1}^{y}S_{2l}^{y}+\Delta S_{2l-1}^{z}S_{2l}^{z}) \Big]  \nonumber \\
& +& \sum_{l=1}^{2N-1} \Big[ (1+(-1)^{l-1} \delta)(S_{2l}^{x}S_{2l+1}^{x}+S_{2l}^{y}S_{2l+1}^{y} \nonumber \\
&+&\Delta S_{2l}^{z}S_{2l+1}^{z}) \Big]
\end{eqnarray}
where  $1-j(-\infty \le j \le \infty)$ and $\delta (-1 \le \delta \le 1)$ represent the degree of dimerization and quadrumerization, respectively. The open boundary condition is assumed. The anisotropy parameter is denoted by $\Delta$. In the present work, we concentrate ourselves on the cases $\Delta = 1$ (isotropic case) and 0 ($XY$ case). We also take $\delta \geq 0$ without loss of generality. 

 Let us discuss the ground state of this model in some limiting cases for $\Delta = 1$. First, our model tends to the $S=1$ AFHC for large negative $j$.  Therefore the ground state is the VBS-like Haldane phase or the $S=1$ dimer phase according as $\delta < \delta_c$ or $\delta > \delta_c$ where $\delta_c$ is the critical value which depends on $j$. For $j \rightarrow -\infty$, $\delta_c$ tend to 0.25 \cite{tk,kno} which is the value for the $S=1$ AFHC. In terms of the spin-1/2 language, these phases can be also described as follows. For small $\delta$ (Haldane phase), the spin pairs connected by the $1+\delta$-bonds and  $1-\delta$-bonds form local singlet pairs as schematically shown in Fig. \ref{fig1}(a). For large $\delta$ ($S=1$ dimer phase), the spins connected by $1+\delta$-bonds ($\v{S_{4l+2}}$ and $\v{S_{4l+3}}$) form  singlet pairs strongly. Mediated by the fluctuation within these singlet pairs, the effective antiferromagnetic coupling is induced between the spins $\v{S_{4l+1}}$ and $\v{S_{4l+4}}$, which leads to the 4-spin local singlets as shown in Fig. \ref{fig1}(b). It should be noted that the above picture in spin-1/2 language remain valid even for positive $j$ as far as $j < 1$.

Secondly, we discuss the neighbourhood of $\delta=0$ and $j \simeq 1$. For $\delta = 0$, the singlet pairs reside on the $j$-bonds or the $1 \pm \delta$ bonds according as $j > 1$ or $< 1$ corresponding to the $S=1/2$ dimerized state with oppsite parity. For $j<1$, the dimer configuration is the same as in Fig. \ref{fig1}(a) while it is shown in Fig. \ref{fig1}(c) for $j>1$.

\section{Numerical Results for $\Delta = 1$}

The Hamiltonian is numerically diagonalized to calculate the energy gap $G(N, j, \delta)$ in the open chain for $4N=12,16,20$ and 24 using the Lanczos algorithm. The open chain is used so that the critical point can be easily discerned. For concreteness, let us take the leftmost and rightmost bonds as $j$-bonds. According to the spin structure discussed is $\S$2,  for small values of $\delta$, there remain two $S=1/2$ residual spins at the both ends of the chain leading to the  4-fold quasi-degeneracy as far as $j < 1$. In this case, the energy gap $G(N, j, \delta)$ decreases exponentially with $N$. On the other hand, for large $\delta$,  there remain no residual spins. Therefore the  energy gap $G(N, j, \delta)$ remains finite in the thermodynamic limit. Thus the product $NG(N, j, \delta)$ decreases (increases) with $N$ for $\delta < \delta_c$ ($\delta > \delta_c$). Thanks to  this situation, we can accurately determine the critical point using  the phenomenological renormalization group method. 

If we search for the critical point varying $\delta$ with fixed $j$, the phenomenological renormalization group equation for the finite size critical point $\delta_c(N_1,N_2)$ reads\cite{ptcp}, 
\begin{equation}
N_1G(N_1, j, \delta_c(N_1,N_2))=N_2G(N_2, j, \delta_c(N_1,N_2)).
\end{equation}
We also have the corresponding equation for $j_c(N_1, N_2)$ if the roles of $j$ and  $\delta$ are interchanged. The finite size critical points are obtained as the intersection of $N_1G(N_1, j, \delta)$ and $N_2G(N_2, j, \delta)$ as shown in Fig. \ref{fig5}. Three intersections for $(N_1, N_2) =(12,16), (16,20)$ and (20,24) are represented by double circles.

These values are extrapolated to $N \rightarrow \infty$ assuming that the finite size correction to the critical value is proportional to $(\frac {N_{1}+N_{2}}{2})^{-1/\nu}$ where $\nu$ is the critical exponent of the correlation length or, equivalently, the energy gap. This assumption is based on the finite size scaling hypothesis\cite{ptcp}. At $j=1$ and $\delta=0$, the exponent $\nu$ is known to be $2/3$ with the logarithmic corrections to the pure power-law behavior\cite{cf,be,sf}. Assuming that $\nu$ remains constant over the whole phase boundary, we take $\nu=2/3$ in the extrapolation of the critical point. This assumption is plausible because the system remains $SU(2)$ symmetric over the whole phase boundary. 

To check this assumption numerically, we estimate the value of the exponent $\nu$ from the numerical data using  the following formula of Spronken et al.\cite{sf} ,
\begin{equation}
\nu(N_1,N_2)=\frac {\ln \left[ (\frac{N_2}{N_1}) \left( \frac{\ln N_1}{\ln N_2} \right)^{1/2} \right]}{\ln \{ \frac{N_2 G'(N_2)}{N_1 G'(N_1)} \} }
\end {equation}
which takes  into account the logarithmic corrections due to the $SU(2)$ symmetry of the isotropic Heisenberg model. Here $G'(N)$ denotes the derivative of $G(N, j, \delta)$ with respect to $j$ or $\delta$ at $j=j_c(N_1,N_2)$ or $\delta=\delta_c(N_1,N_2)$.

Fig. \ref{fig6} and Fig. \ref{fig7} show the examples of the extrapolation procedure of the critical points and critical exponent $\nu$ for $\delta = 0.5$. The cross is the critical point $j_{\rm c} \simeq 0.889$ and exponent $\nu \simeq 0.670$ in the thermodynamic limit.
The $\delta$-dependence of $\nu$ is shown in Fig. \ref{fig8}. The horizontal line is $\nu = 2/3$. It is verified that  $\nu \simeq 2/3$ over the whole phase boundary except for the neighbourhood of $j=0$ and $\delta=1$ where the numerical accuracy becomes worse. In this region, however, the phase boundary and the critical exponent can be determined by the perturbation theory with respect to $j$ as follows. 

For small $j$, the spins connected by the $1+\delta$-bonds form strong singlet pairs. Between the spins $\v{S_{4l+1}}$ and $\v{S_{4l+4}}$, the effective coupling $j_{\mbox{eff}}$ is generated mediated by the quantum fluctuation in the singlet pairs on the $1+\delta$-bonds. This is calculated as,
\begin{equation}
\label{jeff}
j_{\mbox{eff}}=\frac{\frac{1}{2}j^2}{1+\delta}
\end{equation}
up to the second order in $j$. When $1-\delta=j_{\mbox{eff}}$, the effective Hamiltonian is the uniform spin-1/2 AFHC which has a gapless ground state. Thus the phase boundary is given by,
\begin{equation}
\frac{1}{2}j^{2}+\delta^{2}=1,
\end {equation}
and the gap exponent $\nu = 2/3$.

The ground state phase diagram of the present model is shown in Fig. \ref{fig9} by the solid line. The perturbation result is shown by the dashed line. The numerical result is consistent with the perturbation theory.

On the other hand, Takano \cite{takano} has derived the gapless line of the present model as $\delta=(1-j)^{1/2}$ using the mapping onto the nonlinear $\sigma$-model. This is in qualitative agreement with our result as shown in Fig.6 by the dash-dotted line.

Each region of the phase diagram is illustrated as follows. When  $j$ is positive and close to unity, the phases inside and outside the solid line are the $S=1/2$ dimer phases with different parity.  The phase boundary is the gapless line where the energy gap disappears due to the competition of dimerization and quadrumerization. When $j$ is negative, the ground state is the Haldane state inside the solid line and $S=1$ dimer state outside the solid line. The $S=1/2$ dimer phase is thus connected with the  Haldane phase continuously inside the solid line even in the presence of quadrumerization. On the other hand, the connection between the opposite parity $S=1/2$ dimer phase and the $S=1$ dimer phase outside the solid line is interrupted by the point $\delta=1$ and $j=0$. At this point the ground state consists of isolated dimers on the $1+\delta$-bonds and free spins on the $4l$-th and $4l+1$-th sites. In the presence of small but finite $j$, however, the spin structure does not depend on the sign of $j$ because the effective coupling $j_{\mbox{eff}}$ is proportional to $j^2$ according to Eq. (\ref{jeff}). Therefore we may conclude that the $S=1/2$ dimer phase with $j > 1$ and the $S=1$ dimer phase also belong to a single phase.

\section{Analytic Results for $\Delta =0$}

For $\Delta=0$ ($XY$-model), our model reduces to the half-filled noninteracting spinless fermion system by the Jordan-Wigner transformation. After the  Fourier transformation, the single particle excitation spectrum is determined by the eigenvalue of the $4 \times 4$ matrix,

\begin{equation}
\left[
\begin{array}{cccc}
0 & \frac{(1-\delta)e^{ik}}{2} & 0 & \frac{je^{-ik}}{2} \\
\\
\frac{(1-\delta)e^{-ik}}{2} & 0 & \frac{je^{ik}}{2} & 0 \\
\\
0 & \frac{je^{-ik}}{2} & 0 & \frac{(1+\delta)e^{ik}}{2} \\
\\
\frac{je^{ik}}{2} & 0 & \frac{(1+\delta)e^{-ik}}{2} & 0 \\
\end{array}
\right] ,
\end {equation}
where $k$ is the momentum of the excitation. The eigenvalue is given by, 
\begin{equation}
\varepsilon=\pm\varepsilon_{\pm}(k) ,
\end{equation}
where
\begin{eqnarray}
\varepsilon_{\pm}(k)&\equiv& \left[\frac{1}{2}\left\{j^{2}+(1+\delta^{2})\right\}\right. \nonumber  \\ &\pm& \left. \sqrt{j^{2}\cos^{2}{2k}+j^{2}\delta^{2}\sin^{2}{2k}+\delta^{2}} \right]^{1/2}.
\end{eqnarray}

In the half filled case, negative branch is filled and the excitation energy is determined by $\varepsilon_{\pm}(k)$. The energy gap $G$ is  determined from the minimum of  $\varepsilon_{\pm}(k)$ as,
\begin{equation}
G = \sqrt{\frac{1}{2}\big[j^{2}+(1+\delta^{2})\big] - \sqrt{j^{2}+\delta^{2}}}.
\end{equation}

Setting $G=0$, the critical line can be exactly calculated as,
\begin{equation}
j^{2}+\delta^{2}=1,
\end {equation}
and the critical exponent $\nu = 1$. The phase boundary is a circle as shown in Fig. \ref{fig9} by the dotted line.

When $j$ is positive, the phase diagrams of the isotropic case  and that of the  $XY$ case look similar. For negative values of $j$, however, these two models behave quite differently. In the isotropic case, the Haldane-like phase extends to $j \rightarrow -\infty$ for small $\delta$ while it end up at $j = -1$ in the $XY$ case. The phase at $j \rightarrow -\infty$ is continuously connected with the $S=1$ dimer phase. This can be understood in the following way. 

The coupling on the $j$-bond can be rewritten as,
\begin{eqnarray} \nonumber
 &&j(S_{2l-1}^{x}S_{2l}^{x}+S_{2l-1}^{y}S_{2l}^{y}+\Delta S_{2l-1}^{z}S_{2l}^{z})  \\ \nonumber
&=& j\v{S}_{2l-1}\v{S}_{2l}+j(\Delta-1) S_{2l-1}^{z}S_{2l}^{z} \\ \nonumber
&=& j\v{S}_{2l-1}\v{S}_{2l}+\frac{j(\Delta-1)}{2}\{( S_{2l-1}^{z}+S_{2l}^{z})^2-\frac{1}{2}\}.
\end{eqnarray}
If we regard $\v{S}_{2l-1}+\v{S}_{2l}$ as a single $S=1$ spin operator, the last term corresponds to the single site anisotropy term which has large positive value for $j \rightarrow -\infty$ and $0 < \Delta < 1$. Therefore the ground state at $j \rightarrow -\infty$ is the large-$D$ like phase in the $XY$ case. Considering that no phase boundary exists between the large-$D$ phase and the dimer phase in $S=1$ AFHC,\cite{tone} it is reasonable that we find  no phase transition as a function of $\delta$ for large enough negative $j$ in the present model.

\section{Summary and Dicussion}

The ground state phase diagram of the dimerized and quadrumerized spin-1/2 Heisenberg chain is calculated by numerical diagonalization method. In the isotropic case, the critical points are determined using the phenomenological renormalization group and the finite size scaling hypothesis. The gap exponent $\nu$ is estimated to be close to $2/3$ over the whole phase boundary taking the logarithmic corrections into account. It is suggested that the phase transition of the present model belongs to the same universality class as the $S=1/2$ dimerization transition as expected from the symmetry of the system. The numerical results are consistent with the perturbation theory for $\delta \simeq 1$ and $j \simeq 0$. 

It is found that the Haldane phase is connected with the $S=1/2$ dimer phase even in the presence of quadrumerization $\delta$. The $S=1$ dimer phase is found to be connected with the $S=1/2$ dimer phase with opposite parity. At first glance, the spin structures of the last two states might appear different. The difference of these two spin structures is, however, local and can be transformed via a local reconstruction of 4-spin singlet without closing the energy gap. Therefore these two states are continuously connected with each other. Actually, no evidence of the phase transition is observed in the numerical results.

In the isotropic case, our problem has always the $SU(2)$ symmetry and this fixes the value of the critical exponents. If we violate the $SU(2)$ symmetry by introducing the anisotropic exchange interaction, we may expect wider variety of critical phenomena. Considering the big difference between the phase diagrams of the $XY$ case and the isotropic case, it must be of interest to investigate the intermediate region $0 < \Delta <1$ in detail. The study of this problem is in progress and will be reported elsewhere.

Another way of breaking the $SU(2)$ symmetry is to introduce the magnetic field. The effect of the magnetic field is of special interest related with the magnetization plateau which can be regarded  as the field induced Haldane gap problem\cite{yoa,totsuka,tone}. This is also under investigation and will be reported elsewhere.

We thank H. Nakano and K.Takano for useful discussion and fruitful comments. We are also grateful to H. Nishimori for the program package  TITPACK version 2 for the diagonalization of spin-1/2 systems. The numerical calculation is performed using the HITAC S820 at the Information Processing Center of Saitama University and the FACOM VPP500 at the Supercomputer Center of Institute for Solid State Physics, University of Tokyo.

\newpage
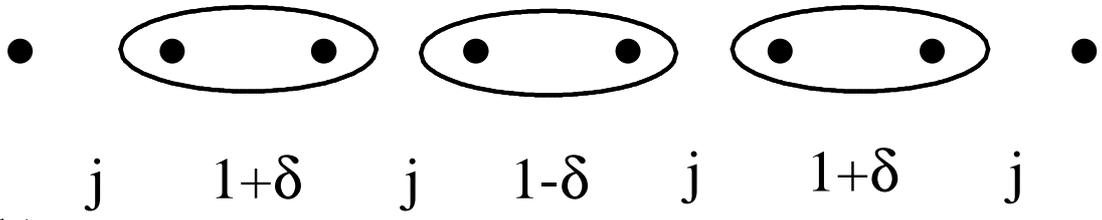
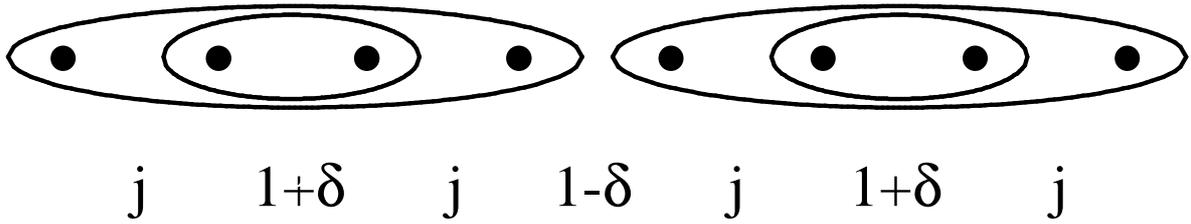
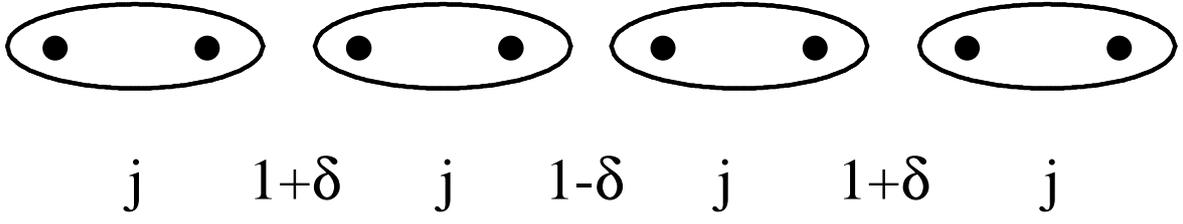
\begin{figure}
\caption{The schematic spin structure for (a) $j < 1$, $\delta < \delta_{c}$, (b)  $j < 1$, $\delta > \delta_{c}$ and (c) $\delta=0$, $j > 1$ }
\label{fig1}
\end{figure}
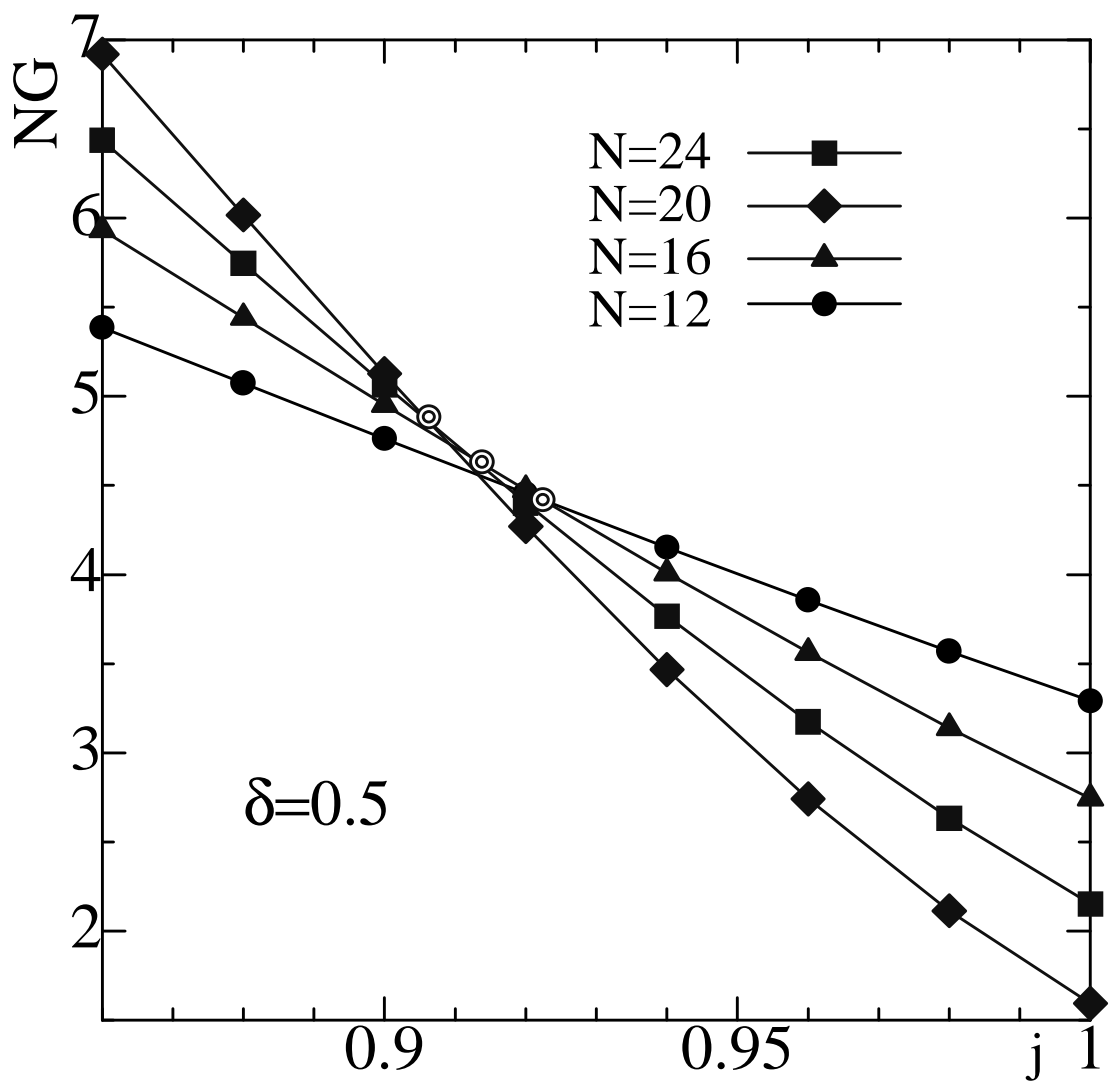
\begin{figure}
\caption{The $j$-dependence of $NG(N,j,\delta)$ with $\delta=0.5$ for $4N=12, 16, 20$ and 24. The intersections (double circles) are the finite size critical points.}
\label{fig5}
\end{figure}
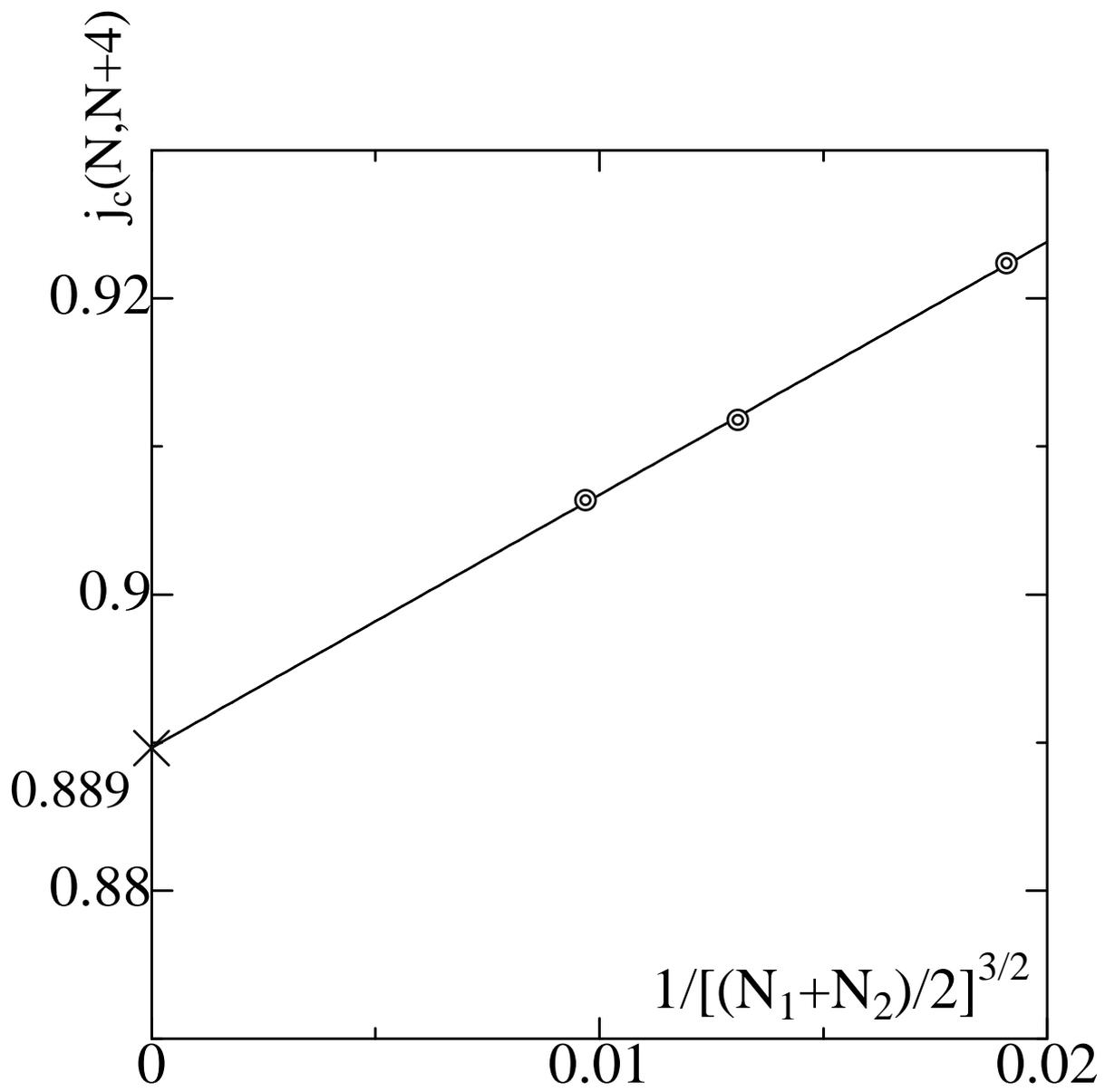
\begin{figure}
\caption{The extrapolation procedure of finite size critical point $j_c$ for $\delta=0.5$}
\label{fig6}
\end{figure}
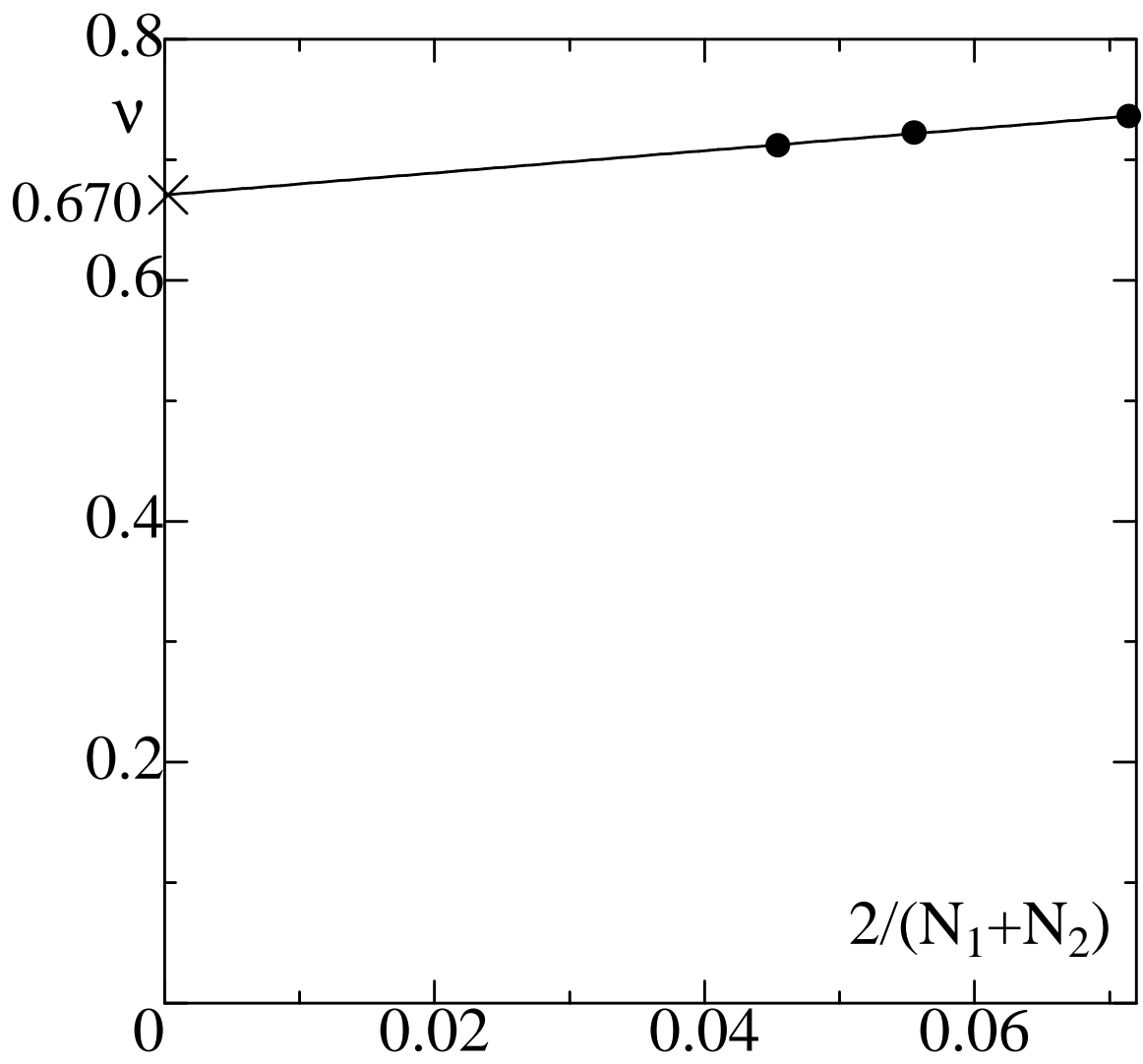
\begin{figure}
\caption{The extrapolation procedure of finite size critical exponent $\nu$ for $\delta=0.5$}
\label{fig7}
\end{figure}
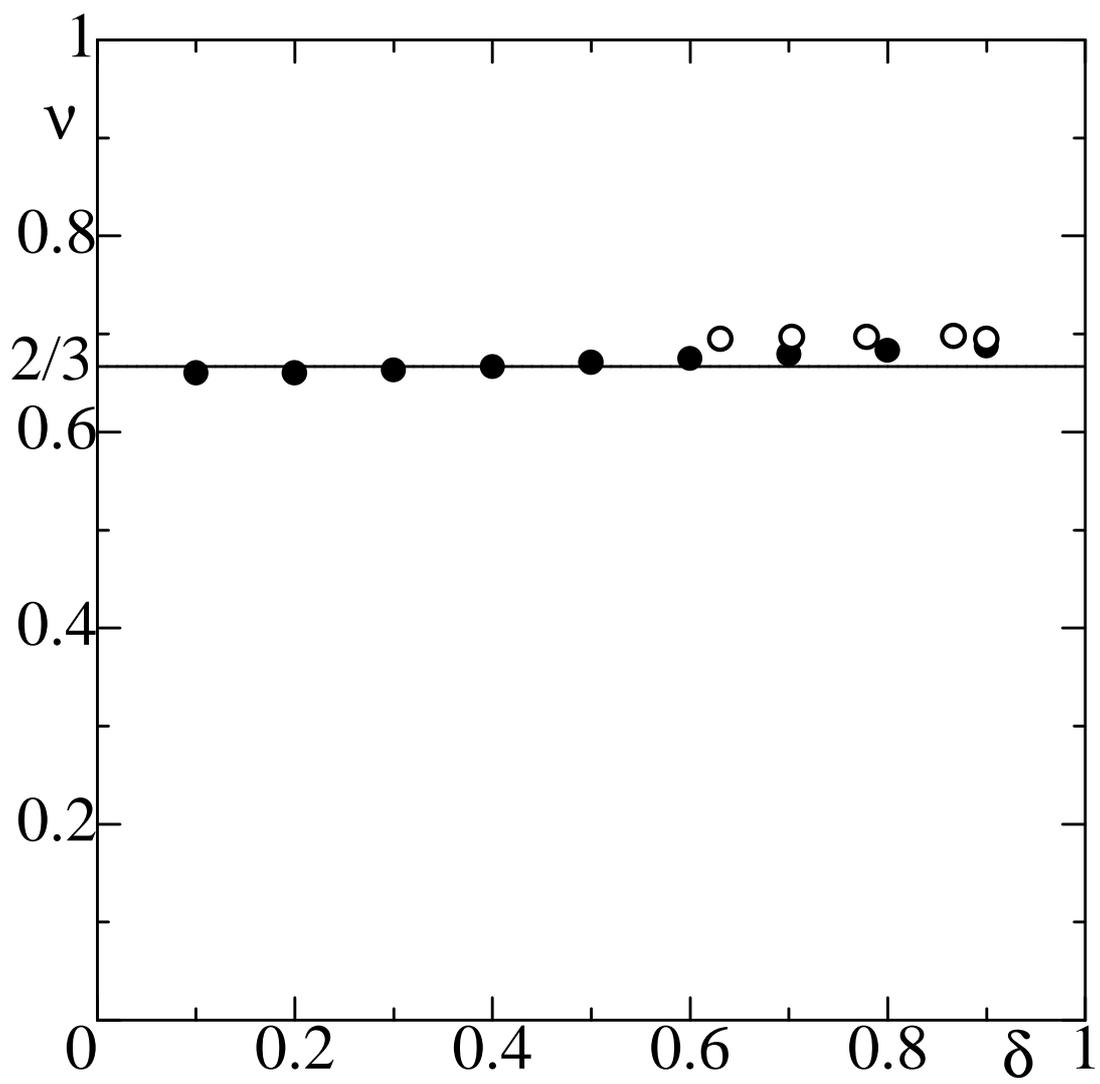
\begin{figure}
\caption{The $\delta$ dependence of the numerically obtained critical exponent $\nu$. Filled (open) circles represent the values for $j>0 (j<0)$.}
\label{fig8}
\end{figure}
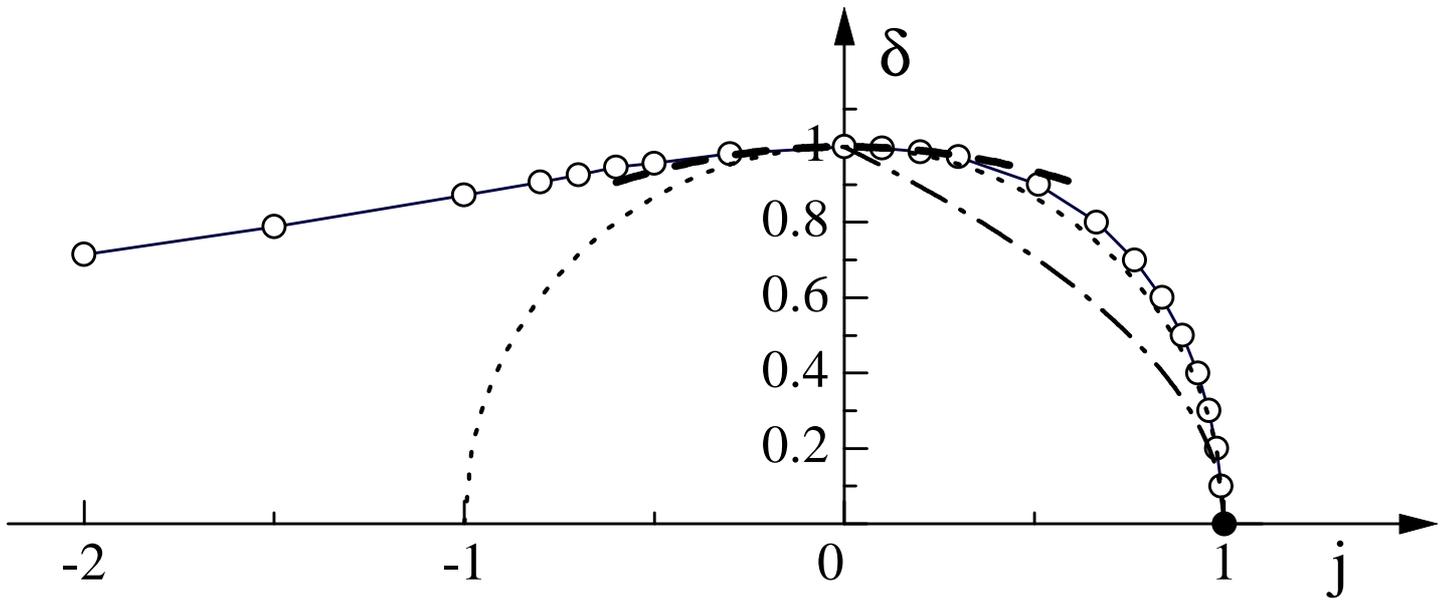
\begin{figure}
\caption{The phase diagram of the isotropic model (solid line) and the $XY$ model (dotted line). The dashed line is the result of the perturbation calculation for $\delta \simeq 1$ and $j \simeq 0$. The dash-dotted line is the phase boundary obtained from the nonlinear $\sigma$ model \cite{takano}.}
\label{fig9}
\end{figure}
\end{document}